\documentclass{aa}

\newcommand {\mc} {\,$\mu$m} 
\newcommand {\cmu} {\,cm$^{-1}$}
\usepackage{graphicx}
\usepackage{txfonts}
\begin{document}
   \title{An IR study of pure and ion irradiated frozen formamide}

      \author{ J. R. Brucato, \inst{1}
                    G.A. Baratta,  \inst{2}\and G. Strazzulla \inst{2}
          }
   \offprints{ J.R. Brucato}

   \institute{INAF Osservatorio Astronomico di Capodimonte, via Moiariello 16,
		80131 Napoli, Italy\\     
		\email{brucato@na.astro.it}             
         \and
 INAF--Osservatorio Astrofisico di Catania, via Santa Sofia 78, I-95123 Catania, Italy\\
             }

   \date{Received  27 February 2006 / Accepted  10 April 2006}
   
   \abstract
    {The chemical evolution of  formamide (HCONH$_2$),  a
  molecule of astrobiological interest that has been tentatively identified in interstellar ices and in
  cometary coma, has been studied in laboratory under simulated astrophysical conditions  such as ion irradiation at low temperature.}
    {To evaluate the abundances of formamide observed in space or in laboratory, the integrated absorbances for all the principal IR features of frozen
  amorphous pure formamide deposited at 20 K were measured. Further evidence that energetic processing of ices occurring in space is extremely relevant both to astrochemistry and to astrobiology has been found, showing that new molecular species are synthesized by ion irradiation at a low temperature.}
     {Pure formamide were deposited at 20 K and IR transmission spectra measured for
  different ice thicknesses. The ice thickness was derived by looking at the interference pattern
  (intensity versus time) of a He-Ne laser beam reflected at an angle of 45 deg by the vacuum-film
  and film-substrate interfaces. Samples of formamide ice were irradiated with 200 keV H$^+$ ions
  and IR spectra recorded at different ion fluences.} 
    {New molecules were synthesized among which are CO, CO$_2$, N$_2$O, isocyanic acid (HNCO),
  and ammonium cyanate (NH$_4$$^+$OCN$^-$).  Some of these species remain stable after warming
  up to room temperature.}
  {}
 
  \keywords{astrochemistry --
  	   astrobiology --
            methods: laboratory --
            techniques: spectroscopic --
            ISM: molecules --
            Comets: general
            }

 \titlerunning{Irradiation of frozen formamide}
 \authorrunning{J.R. Brucato \& al}
  
   \maketitle

\section{Introduction}

Dust particles are mostly formed in circumstellar regions of evolved stars and ejected in the interstellar medium. In dense molecular clouds, most of the gas--phase species are rapidly condensed on refractory cores, forming icy mantles. The frozen mantles are rich in water, the dominant ice in the Universe, which was identified as a component of dust grains (Willner et al. 1982). Space--based infrared observations performed by observatories such as ISO or Spitzer show that new icy species are continuously detected.

Formamide (HCONH$_2$) was observed in the interstellar medium (for a review of molecules observed in interstellar and circumstellar media, see Millar 2004), in the long period comet C/1995 O1 Hale--Bopp (Bockele\'e-Morvan 2000), and tentatively in young stellar objects W33A (Schutte et al. 1999) and NGC 7538 IRS9 (Raunier et al 2004). Formamide is very interesting for its active role in prebiotic chemistry. The chemical reactions of simple compounds containing H, C, N, and O, such as formamide,  are considered a plausible pathway for synthesis on the Earth of biomolecules under prebiotic conditions (Oparin 1938, Miller 1953, Eschenmoser \& Loewenthal 1992, Chyba \& McDonald 1995, Saladino et al. 2004). Among the different theories of prebiotic chemistry active in the early Earth (Miller--Urey and Fischer--Tropsh synthesis), the extraterrestrial synthesis of organic compounds and delivery  on the planetary surface is an interesting one. A considerable amount of extraterrestrial material was and is continuously delivered on the Earth.  It was estimated that between 10$^7$ and 10$^9$ kg yr$^{-1}$ of carbon contained in organic compounds arrived here in the first billion years as interplanetary dust particles (IDPs) (Chyba \& Sagan 1992). Moreover, different strategies for synthesizing organic molecules occurring in space consider simple nitrogen--bearing compounds -- such as hydrogen cyanide (HCN), isocyanic acid (HNCO), formamide, ammonium cyanide (NH$_4$CN), or mixtures of H$_2$O, CO$_2$, CO,  NH$_3$,CH$_4$, CH$_2$O, CH$_3$OH, etc. -- as potential astrobiological precursors (Grim \& Greenberg 1987, Demyk et al. 1998, Moore \& Hudson 2003, Hudson et al. 2001, Palumbo et al 2000, 2004). In this work we focused on pure formamide molecule under simulated astrophysical conditions. 

The HCONH$_2$ molecule is formed at room temperature by the hydrolysis of HCN, and it is the most abundant product of the pyrolysis of HCN--polymer. It is a reactive compound  both at the carboxyl moiety and at the amino group. The role of formamide as prebiotic precursor of the synthesis of nucleic acid bases has been shown under a variety of conditions. In particular the presence of different inorganic compounds, such as oxides, minerals or cosmic dust analogues, is able to catalyze the formamide condensation that originate, among many compounds,  purine and pyrimidine  bases, namely, the products of the first primordial steps that might have leaded to the appearance of life as we know it (Yamada et al. 1972, 1975, Saladino et al. 2003, 2004, 2005). 

Energetic processing of ices occurring in space is extremely relevant both to astrochemistry and to astrobiology.  Processing by UV photons of interstellar/cometary ice analogue mixtures of  H$_2$O, CH$_3$OH, CO, and NH$_3$ has shown that during warming up to 200 K formamide, acetamide, ethanol, and nitriles compounds are formed (Bernstein et al. 1995). Moreover, it was suggested that formamide is one of the products of the ion irradiation and phoyolysis of H$_2$O + HCN (Gerakines et al. 2004) or of NH$_3$ + CO (Demyk et al 1998), even at a low temperature (18 K). Unfortunately, a quantitative measurement of the amount synthesized has not been made due to lacking IR band absorbances. Experimental investigations of chemical evolution of pure formamide in simulated interstellar/cometary conditions are therefore deemed necessary.

In this paper we present an IR transmittance spectroscopy study in the 4000--1200 \cmu range of pure formamide  deposited at a low temperature (20 K). We measured the absorbances of the major IR bands and irradiated samples of condensed formamide with 200 keV H$^+$ ions at different fluences. New molecules were synthesized and identified in the spectra. The chemical evolution of the irradiated samples were then monitored at increasing temperatures. 


\begin{table}
\caption{ Peak positions, vibration assignments, and integrated absorbances of  frozen (20 K) amorphous formamide.}
\label{table1}
\centering
\begin{tabular*}{\linewidth}{l@{\extracolsep{\fill}}*{4}{l}} 
\hline
\noalign{\smallskip}
\multicolumn{3}{c}{Peak position} & & Integrated absorbance \\ 
\cmu & &\mc  & Assignment & ($\times$10$^{-17}$cm~mol$^{-1}$) \\ 
\noalign{\smallskip}
\hline
\noalign{\smallskip} 
3368  && 2.97 &  {$\nu$}{$_{1}$}  asym. NH$_2$   stretch  & 13.49 ({$\nu$}{$_{1}$}+{$\nu$}{$_{2}$}) \\
3181  && 3.14 &  {$\nu$}{$_{2}$}  sym. NH$_2$   stretch \\
2881  && 3.47 &  {$\nu$}{$_{3}$}  CH   stretch  & 0.47\\
1708  && 5.85 &  {$\nu$}{$_{4}$}  CO stretch & 6.54 ({$\nu$}{$_{4}$}+{$\nu$}{$_{5}$})\\
1631  && 6.13 &  {$\nu$}{$_{5}$}  in plane NH$_2$   scissoring \\
1388  && 7.20 &  {$\nu$}{$_{6}$}  in plane CH   scissoring & 0.68\\
1328  && 7.53 &  {$\nu$}{$_{7}$}  CN stretch & 0.85 \\
\noalign{\smallskip}
\hline
\end{tabular*}
\end{table}

\section{Experimental procedures}

Gaseous samples of formamide were prepared in a pre--chamber (P $<$ 10$^{-6}$~mbar) and admitted by a gas inlet through a needle valve into the scattering chamber, where they accreted onto a cold (20 K) silicon substrate; the  thickness of the deposited films were then measured (see below for details). The infrared transmittance spectra were obtained  in the high--vacuum scattering chamber  (P $<$ 10$^{-7}$~mbar) interfaced with a FTIR spectrophotometer (Bruker Equinox~55) through IR--transparent windows. The silicon substrate was in thermal contact with a closed--cycle helium cryostat whose temperature can be varied between 10 and 300 K. The vacuum chamber is interfaced with an ion implanter (200~kV; Danfysik 1080--200) that generates ions with energies up to 200~keV (400~keV for double ionization). The ion beam produces a spot that is larger than the area probed by the infrared beam (for more details on the experimental set up see Strazzulla et al. 2001).
In this work we used 200 keV H$^+$ ions.
In order to avoid macroscopic heating of the target, the current density was maintained in the range of 100~$\mathrm{nA\, cm^{-2}}$ to a few  $\mathrm{\mu A\, cm^{-2}}$.
The ion fluence in $\mathrm{ions\, cm^{-2}}$ was
measured by a charge integrator from the ion current monitored during irradiation.
The substrate plane is placed at an angle of 45~degrees with respect to the IR beam and the ion beam so that spectra can be taken in situ, even during irradiation, without tilting the sample. Spectra were taken at selected temperatures in the range of 20--300 K. All the spectra shown below were taken with a resolution of 1~cm$^{-1}$ using a DTGS detector.

The 200 keV H$^+$ ions penetrate about 2~{$\mu$m} in formamide calculated using the TRIM program (Transport of Ions in Matter; e.g., Ziegler 1977;  Ziegler et al. 1996). The maximum thickness of deposited layers was about 0.49~{$\mu$m} (see below) i.e. thinner than the penetration depth of incoming ions. With the same software, it is also possible to calculate the stopping power (i.e. the amount of energy deposited per unit path length) of a given ion in a given target. For 200 keV H$^+$ in formamide we obtained about 50$\times$10$^{-15}$ eV cm$^{2}$/molecula.  By multiplying this number times the number of bombarding ions per square centimetre we obtained the amount of energy released to the sample (dose) in eV per molecula. Here we express the dose in eV/16amu, because it is a convenient way to characterize chemical changes and to compare with other experiments with different samples.

   \begin{figure}
   \centering
   \includegraphics[width=10cm]{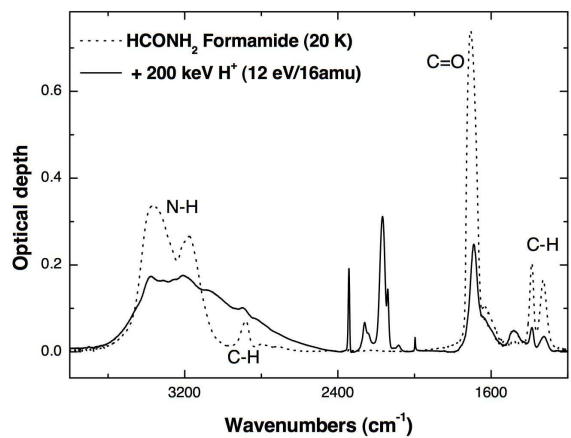}
      \caption{IR transmittance spectrum  in optical depth scale of formamide as deposited at 20 K on a silicon substrate (dot line). The band--peak positions and assignment are given in Table 1. The spectrum obtained after irradiation at a dose of 12 eV/16amu with 200 keV H$^+$ ions is also shown (full line).}
         \label{fig1}
   \end{figure}

\section{Results}

Figure~\ref{fig1} shows the two spectra of (HCONH$_2$)  before  and after  ion irradiation (12 eV/16 amu deposited by 200 keV H$^+$ ions)  at 20 K in the 3800--1200 \cmu range. The peak positions of fundamental vibrational modes of formamide and their assignments  are reported in Table~\ref{table1}. Such assignments were made on the basis of work on gaseous or liquid samples (McNaughton et al. 1999, Rubalcava 1956). The (HCONH$_2$) spectrum is characterized by a couple of intense and broad bands at 3368 and 3181 \cmu, which correspond to asymmetric and symmetric NH$_2$ stretching modes, respectively. The strongest band is peaked at 1708 \cmu and related to the CO--stretching mode of the carbonyl group.

\subsection{Absorbances}

The experimental set--up allows the thickness to be monitored during accretion by looking at the interference pattern (intensity versus time) given by a He--Ne laser beam reflected at an angle of 45 deg by the vacuum--film and film--substrate interfaces. After the reflection from the substrate, the laser beam follows the same path of the infrared beam coming from the IR source and can be detected by using an external detector placed in the source compartment of the IR spectrometer. 

In general the interference curve versus thickness given by the laser beam reflected by the film+substrate assembly is an oscillating function. For absorbing materials, the laser light transmitted into the film and reflected back by the interface film--substrate is attenuated into the film; thus the amplitude of the oscillation in the interference curve exponentially decays with the thickness, and the reflectance approaches its bulk value at a large thickness. In molecular ices at visible wavelengths, the absorption is so low that it can be neglected for a thickness of a few microns, so that the interference curve can be considered a periodic function whose period (distance between two maxima or minima) is given by the relation: 
\begin{equation}
\Delta d=\frac{\lambda _{0}}{2n_{f}\sqrt{1-sin^2\theta_i/n^2_f}}
\label{eq:inter}
\end{equation}
where $n_{f}$ is the refractive index of the ice at the laser wavelength 
$\lambda_{0}$, and $\theta_{i}$ is the incidence angle measured from the normal. The previous relation can be used to measure the thickness if $n_{f}$ is known. The amplitude of the interference curve depends on the refractive index of the ice, the refractive index of the substrate, the incidence angle, and on the polarization of the laser. Hence $n_{f}$ can be derived from the amplitude of the interference curve (intensity ratio between maxima and minima); details of this method can be found in Westley et al (1998) and Baratta and Palumbo (1998). 

The absolute accuracy of the thickness measured in this way is about 5$\%$ and is limited mainly by the uncertainties in the knowledge of the refractive index of the substrate at low temperature (silicon in our case) and by the error in measuring the incidence angle of the laser. By following this method we found a refractive index $n_{f}$=1.361 for formamide ice at the laser wavelength of 0.543 \mc. For an incidence angle of 45 deg, this yields a film thickness of 0.234 \mc~between interference maxima. Once the thickness is measured, the integrated bands absorbance, {\it A} (cm mol$^{-1}$), can be computed from the infrared spectrum if the density $\rho$ of the film is known through the formula:
\begin{equation}
A=\frac{\int \tau(\nu)d\nu}{N}
\end{equation}
where $\tau(\nu)$ is the optical depth and $N$ the molecular column density in units of molecules cm$^{-2}$.

In this work we derived the density of formamide ice by using the Lorentz--Lorenz relation:
\begin{equation}
L \rho=\frac{n_{f}^{2}-1}{n_{f}^{2}+2}.
\label{eq:inter}
\end{equation}
For a given material, the quantity $L$ (Lorentz--Lorenz coefficent) is nearly constant for a given wavelength regardless of the material phase and temperature (Wood \& Roux 1982). Formamide at 25 C (liquid) has a density $\rho^{liq}$=1.129 gcm$^{-3}$ and a refractive index of $n^{liq}_{f}$=1.446 for the sodium line (0.589 \mc) (Cases et al. 2001). The corresponding Lorentz--Lorenz coefficient is $L$= 0.2362~cm$^{3}$~g$^{-1}$. By substituting in Eq~(\ref{eq:inter}), the L coefficient derived for the liquid phase and the refractive index measured by interference, we obtain a density $\rho^{ice}$=0.937~g~cm$^{-3}$ for formamide ice.
The corresponding integrated band absorbances are given in Table~\ref{table1},  together with band peak positions and assignments. The absorbances were derived by considering two different thickness increased by a factor (from the Snell's low) of $1/cos\theta_{r}=1/\sqrt{1-sin^{2}\theta_{i}/n^{2}_{f}}$, where $\theta_{r}$ is the refractive angle. The correction takes into account, in a approximate way, the increased path length of the IR beam at an oblique incidence of $\theta_{i}$=45~deg. The corrective factor was derived by assuming a constant value of the refractive index with the wavelength. This approximation neglects the variation of the refractive index in the infrared region due to the vibrations and the contribution of the electronic transitions to the dispersion from the visible (0.543~\mc~) to the infrared. We estimate a corresponding uncertainty of $\simeq$~10~$\%$ in the band absorbances.

\subsection{Ion--induced synthesis of molecules}
Figure~\ref{fig1} also shows the IR spectrum of formamide after 200 keV proton irradiation at 12 eV/16amu dose. The peak positions and identification of each new molecular species is reported in Table~\ref{table2}.

\begin{figure}
     \includegraphics[width=10cm]{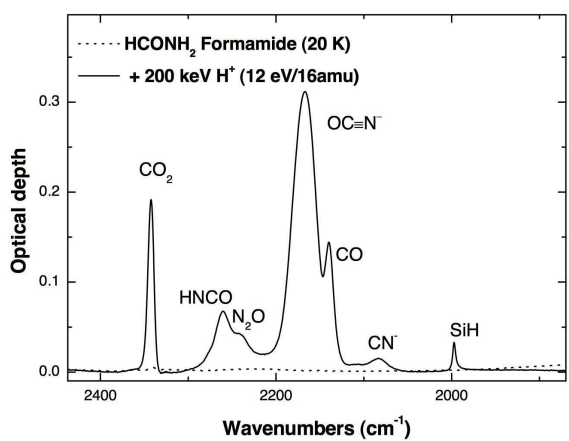}
      \caption{IR transmittance spectrum in optical depth scale of formamide (20 K) after irradiation at a dose of 12 eV/16amu with 200 keV H$^+$ ions (full line). The spectrum obtained before irradiation  is also shown (dotted line).}
         \label{fig2}
   \end{figure}

In the 2430--1850 \cmu spectral region shown better in Fig.~\ref{fig2}, both CO and CO$_2$ absorption bands are observed at 2140 and 2342 \cmu, respectively. The band observed at 2260 \cmu is assigned to the NCO stretching mode of isocyanic acid (HNCO). The band peaked at wavenumbers larger than that observed by Raunier et al. (2004) (2252 \cmu). This may be due to the presence of a shoulder at 2238 \cmu attributed to the synthesis of NO$_2$. The weak band at 2083 \cmu is due to the cyanate anion CN$^-$ stretching mode.  A shoulder at about 2100 \cmu is present  and tentatively assigned to hydrogen cyanide (HCN). The presence of the SiH stretching band observed at 1997 \cmu stems from the fraction of protons that entirely cross the ice sample and is implanted into the silicon substrate.

Another intense band located at 2165 \cmu is commonly assigned to cyanate anion OCN$^-$ (Grim and Greenberg 1987, Hudson et al. 2001, Broekhuizen et al. 2004), even if other carriers have been proposed by Pendleton et al. (1999). Evidence that the ammonium cyanate complex NH$_4^+$OCN$^-$ is formed is given by the wide band at 1478 \cmu, which is ascribed to NH$_4^+$ (Raunier et al. 2003). Further evidence of ammonium ion synthesis is formed in the smooth and weak peaks at 3074 and 3206 \cmu 
observed over the very broad feature extending from about 3600 \cmu to 2400 \cmu . This feature is due to the overlapping of a number of NH and CH stretches of newly formed compounds and of a residual amount of formamide still present after ion irradiation. A tentative identification of NH$_3$ molecules is made by the band  present at 3376 \cmu due to NH vibration stretch and by a noisy feature at about 1110 \cmu that is not reported in Fig.~\ref{fig1}.

   \begin{table}
      \caption{Peak positions, vibration modes, and molecule assignments of newly formed IR bands  after irradiation of formamide at 20 K with 200 keV H$^+$ ions.}
         \label{table2}
      \centering
         \begin{tabular*}{\linewidth}{l@{\extracolsep{\fill}}*{4}{l}} 
            \hline
            \noalign{\smallskip}
\multicolumn{2}{c}{ Peak position} & &
\\
{\cmu}& {\mc}  & Vibration & Assignment 
\\ 
\noalign{\smallskip}
            \hline
            \noalign{\smallskip} 

3376&  2.96  & N--H stretch  & NH$_3$
\\
3206&   3.12  & N--H stretch  & NH$_4^+$
\\
3074 &  3.25  & N--H stretch  & NH$_4^+$
\\
2342  & 4.27 &  C=O stretch & CO$_2$    
\\
2260  & 4.42 &  N=C=O$_{asym.}$ stretch & HNCO 
\\
2238  & 4.47 &  N=N   stretch & N$_2$O    
\\
2165  & 4.62 & N=C=O$_{asym.}$ stretch & OCN$^-$
\\
2140  & 4.67 &  C$\equiv$O stretch & CO
\\
2083  & 4.80 &  C$\equiv$N stretch & CN$^-$ 
\\
1997 & 5.01 &  Si--H stretch & SiH
\\
1478  & 6.77 &  N--H$_{sym.}$ bending & NH$_4^+$   
\\
\noalign{\smallskip}
\hline
         \end{tabular*}
  \end{table}
%

\subsection{Temperature effects}

\begin{figure}[htbp]
   \includegraphics[width=9cm]{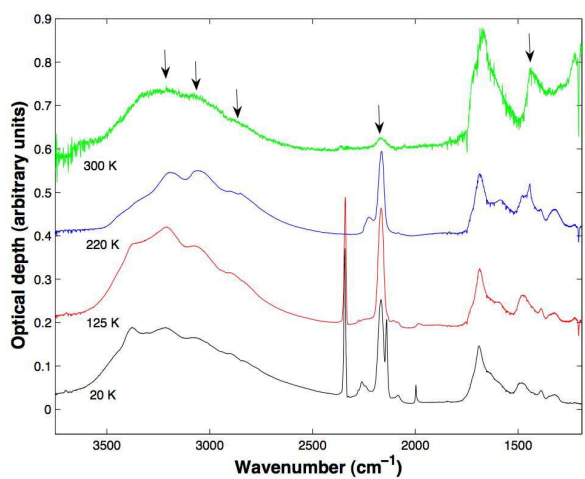}
   \caption{Comparison of the spectra of (from the bottom to the top): formamide  after irradiation with a dose of 24 eV/16amu at 20 K, and after warming up at 125 K, 220 K and room temperature (this spectrum was magnified 5 times). The arrows show the bands corresponding to the presence of  NH$_4^+$OCN$^-$.  Spectra have been arbitrarily shifted for sake of clarity.} 
         \label{fig3}
  \end{figure}

After irradiation, the samples were warmed up (about 1 K/min) to study the evolution of the IR spectrum with temperature, a circumstance that induces a differential sublimation of volatiles and an increase in the synthesized yields likely to be affected by higher mobility. In Fig.~\ref{fig3}  the IR spectra in the range 3700--1200 {\cmu} of formamide after irradiation at a dose of 24 eV/16amu at 20 K and after warming up at 125 K, 220 K, and 300 K are shown. Spectra were arbitrarily shifted and, moreover, that at room temperature magnified five times for sake of clarity.

The peak position of OCN$^-$ shifts during warm--up from 2167 {\cmu} at 20 K to 2164 {\cmu} at 220 K. At room temperature, the OCN$^-$ peak position is about 2167 {\cmu}. The peak position dependence on temperature of OCN$^-$ synthesized by ion irradiation of C,N,O--bearing ice mixtures is a commonly observed phenomena. Nevertheless the amount of shift and the trend observed with temperature depend strongly on the specific ice mixture (Palumbo et al. 2000, Palumbo et al. 2004). In particular the peak position of the OCN$^-$ feature observed in the organic residues at room temperature can vary from 2150 to 2168 {\cmu} depending on the particular ion--irradiated ice mixture considered (Palumbo et al. 2004).
A shift at lower wave number is observed for the N--H$_{sym.}$  bending band peaking at 1437 {\cmu} at 300 K.

It is evident that the volatile species are desorbed and the bands of ammonium cyanate are left over. The formation of NH$_4^+$OCN$^-$ agrees with that measured by the reaction at 10 K of co--deposited NH$_3$ and HNCO (Raunier et al. 2004). The band strength of ammonium cyanate increases with temperature, which could be due to additional formation by reaction between NH$_3$ and HNCO synthesized at low T. Further features observed in Fig.~\ref{fig3}  at 220 K correspond to crystalline formamide still present in the sample after irradiation, which has had a structural transition at about 180 K.

 \section{Discussion}

Formamide has been tentatively identified in the ISO--SWS infrared spectra of different astronomical environments. Although it has been synthesized by UV photolysis and proton irradiation of HCN containing ices (Gerakines et al. 2004) or by UV irradiation of HNCO ice (Raunier et al. 2004) and H$_2$O, CO, NH$_3$ ice mixtures (Demyk et al. 1998, Broekhuizen et al. 2004), it has not been possible to precisely determine the amount of synthesized (or observed) formamide because of the lack of measured integrated absorbances. This work covers such a need by measuring the integrated absorbances for all the principal features of frozen amorphous pure formamide deposited at 20 K.

With the hypothesis that solid formamide is present in protostellar sources as might be the case for NGC 7538 IRS9 (Raunier et al. 2004), the experimental results reported here on the ion processing of formamide ice have to be considered when astronomical and laboratory spectra are compared.
In fact, the evolution of solids (ices, silicates, and carbonaceous phases) in the circumstellar and interstellar media is governed by a number  of processes, such as surface  reactions, UV photolysis,  particle irradiation, and thermal annealing. 

Irradiation in the circumstellar environments of young stars or in the dense interstellar medium was simulated by irradiating pure formamide layers at 20 K. After irradiation, warming the target up causes the sublimation of volatile molecules. This simulates the thermal processing suffered by the dust near a forming star or the  passage of the dust from a dense to a diffuse medium. It is important to note that in this case, calculations indicate that  in the diffuse interstellar medium the temperature 
of the dust is still of the order of 10 K, as it is in the dense medium (Greenberg 1971; Mathis et al. 1983); however, the gas density is such that icy mantles cannot  be maintained or formed.
 
The flux of low--energy cosmic rays irradiating grains in the
interstellar medium is not well known, although a reasonable estimate was given by Moore (1999). She calculated that in cold dense clouds, ions deposit about 30 eV/molecule (in 10$^8$ years), a factor of 10 less than in the diffuse medium. The dominant contribution of cosmic rays comes mainly from low--energy protons in the MeV range, which lose energy  through ionizations and excitations of the target atoms as 200 keV H$^+$ ions do. 
From the present experimental results we estimated that  64 and 78\% of the formamide molecules are destroyed
at 12 and  24 eV/16 amu, respectively.  This result indicates that the destruction of formamide is a first--order process. Extrapolating to the irradiation dose expected in the dense medium, it is possible to infer that, in first approximation,  about 20\% of the frozen formamide molecules are able to survive in dense medium on a time scale of 10$^8$ years.

A very important result is that the species produced after irradiation of formamide are mostly the same as those produced in a large number of irradiation experiments conducted by different groups on different icy mixtures containing simple H,O, C, and N bearing molecules (Grim et al., 1987, Demyk et al. 1998, Moore \& Hudson 2003, Hudson et al. 2001, Palumbo et al 2000, 2004). Moreover, most of the bands shown in Fig.~\ref{fig2} have been observed in astronomical spectra (Sandford et al. 1990, Elsila et al. 1997, Novozamsky et al. 2001,Pendleton et al. 1999, Tegler et al. 1993, 1995, van Broekhiuzen 2005). Of particular interest is the synthesis of the ammonium cyanate molecule. In fact the CN stretching in OCN$^-$ is considered  responsible for the 2165 \cmu band observed in almost all the observations of molecular clouds and in young stellar objects. Ion and UV processing of ice samples containing N--bearing molecules such as N$_2$, NH$_3$, HCN, and HNCO mixed with the most abundant molecules observed in space -- H$_2$O, CO, CO$_2$, and CH$_3$OH -- share the yielding of NH$_4$OCN$^-$ with the results of  this work. 

Ammonium cyanate is theoretically very important since the first synthetic 
production of an organic from inorganic compounds (Wohler 1828). Once prepared by the gaseous reaction of ammonia and cyanic acid, thermal annealing of ammonium cyanate aqueous solution formed urea (NH$_2$)$_2$CO through dissociation into ammonia and isocyanic acid   (Warner and Stitt 1933). Urea was also synthesized at 10 K by UV irradiation of isocyanic acid  and tentatively detected in a protostellar object (Raunier et al. 2004). 

These results suggest that the chemistry of HCN (in the presence of H$_2$O) has to be considered a preferential route for the prebiotic chemistry even in space. 

The plausible pathways for the synthesis of biomolecules in space at low temperature have to be deal with when studying the organic compounds that might have favoured the emergence of life when delivered on the early Earth by comets, meteorites and IDPs. Ion and photon processing of ices containing H, C, N, and O shows a specificity in driving the chemistry in space so that it requires  deep investigation, in particular for its important consequences on astrobiological research. 

\begin{acknowledgements}
This research has been supported by the Italian Space Agency (ASI), and the Italian Ministero dell'Istruzione, Universit\` a e Ricerca (MIUR).
\end{acknowledgements}


\bibliographystyle{aa} 
\bibliography{referenc} 

\end{document}